\documentclass[openacc]{rstransa}

\usepackage{bm}
\usepackage{booktabs}
\usepackage[normalem]{ulem}

\newcommand{\be}{\begin{equation}}
\newcommand{\ee}{\end{equation}}
\newcommand{\bse}
{\begin{subequations}}
\newcommand{\ese}{\end{subequations}}
\newcommand{\bea}{\begin{eqnarray}}
\newcommand{\eea}{\end{eqnarray}}
\newcommand{\ba}{\begin{array}}
\newcommand{\ea}{\end{array}}
\newcommand{\bi}{\begin{itemize}}
\newcommand{\ei}{\end{itemize}}
\newcommand{\ben}{\begin{enumerate}}
\newcommand{\een}{\end{enumerate}}
\newcommand{\del}{\partial}
\newcommand{\im}{\textrm{i}}

\titlehead{Research}
\begin{document}

\title{Quantum turbulence in the many-body regime}

\author{Sayak Bhattacharjee$^{1}$, Mahendra K. Verma$^{2,3}$,  Alexander V. Balatsky$^{4,5}$ and Srinivas Raghu$^{1}$}
\address{
$^{1}$Leinweber Institute for Theoretical Physics, Stanford University, Stanford, California, 94305, USA\\
$^{2}$Department of Physics, Indian Institute of Technology, Kanpur, India. \\
$^{3}$Kotak School of Sustainability, Indian Institute of Technology, Kanpur, India.\\
$^{4}$Nordita, Stockholm University and KTH Royal Institute of Technology,
Hannes Alfvens vag 12, SE-106 91 Stockholm, Sweden,\\
$^{5}$Department of Physics and Institute for Materials Science,
University of Connecticut, Storrs, Connecticut 06269, USA}

\subject{Quantum mechanics, condensed matter physics, statistical mechanics, fluid mechanics}
\corres{Sayak Bhattacharjee\\
\email{sayakbhattacharjee@stanford.edu}}

\begin{abstract}
We discuss phenomenology associated with turbulent hydrodynamics in quantum fluids from a condensed-matter perspective. We begin with weakly-interacting superfluids, often modeled by a mean-field theory governed by the Gross-Pitaevskii equation. Considering the effect of quantum fluctuations beyond the mean-field approximation, we propose a study of many-body quantum effects in turbulent hydrodynamics, especially near zero temperature. We motivate examples of quantum many-body systems where such effects may be uncovered. These include bosons confined in a periodic potential in low spatial dimensions (one and two), and the associated quantum critical point of the superfluid-insulator transition, realized in present-day ultracold-atom and quantum computing platforms. We conclude by listing a set of (open) questions that may be answered using modern quantum many-body techniques. 

This article is part of the theme issue ‘Frontiers of
turbulence and statistical physics’.
\end{abstract}

\begin{fmtext}
\end{fmtext}
	
\maketitle

\section{Introduction}

A prevalent theme in modern theoretical physics is understanding the emergence of complex behavior from a simple set of organizing principles \cite{Anderson:Science1972, laughlin2008different}. In condensed matter systems close to equilibrium, the set of possible phases of matter have been classified extensively based on Landau’s concept of broken symmetry \cite{landau1937theory}, as well as the Landau-Ginzburg-Wilson paradigm based on the idea of the renormalization group \cite{wilson1975renormalization}. Several phases of matter, especially near zero temperature, require quantum mechanics in their many-particle organization. Ground states of many-body quantum systems, with an energy gap to excitations, have been classified based on Landau’s paradigm, supplemented by notions of topological order \cite{wen2017colloquium, senthil2015symmetry}. While a complete classification of gapless ground states is still lacking, a key organizing principle is provided by Landau's Fermi liquid theory, which gives a successful theory of conventional metals \cite{landau1959theory, pines2018theory}. The behavior \textit{near} equilibrium has also been understood in terms of ground state properties, using linear response theory \cite{martin1968measurements, chaikin1995principles}. 

Analogous understanding of the principles governing emergence that results \textit{away} from equilibrium remains far from complete. A fundamental question in this regard has been of the approach to equilibrium, where typically memory of the initial state gives way to a simplified description in terms of state variables such as temperature and pressure. This idea was formulated for quantum many-body systems in the absence of an external reservoir (isolated systems), where it was understood that the system thermalizes as its subsystems act as effective reservoirs for one another \cite{srednicki1994chaos, deutsch1991quantum, nandkishore2015many, d2016quantum}. Of equal interest has been in universal dynamical behavior of many-particle systems as they equilibrate \cite{polkovnikov2011colloquium}. In both classical and quantum many-body systems, the study of hydrodynamic descriptions of the collective dynamics \cite{landau1987fluid, doyon2025generalized}, as well as coarsening behavior \cite{bray1994theory, samajdar2024quantum, gazo2025universal, rodriguez2021turbulent, rodriguez2022far, manovitz2025quantum, andersen2025thermalization}, have had significant impact on these questions. Developments in this regard have routinely benefited from experiments in ultracold atom setups \cite{polkovnikov2011colloquium, bloch2012quantum}, accessing dynamical regimes that are still unreachable in quantum materials. 

Our interest in this article is in the hydrodynamical behavior of quantum many-body systems, when probed \textit{far} from equilibrium\footnote{In the theory of hydrodynamics, one typically assumes that each fluid cell has already achieved local thermodynamic equilibrium. By 'far from equilibrium', we thus refer to a non-equilibrium steady-state achieved by the fluid even within such a description, as in fully-developed turbulence (defined ahead).}. Hydrodynamics represents a coarse-grained description of physical systems in terms of their global conserved quantities, and other slow degrees of freedom such as long wavelength order parameter fluctuations and Nambu-Goldstone modes \cite{chaikin1995principles}. In the case of systems near equilibrium, one expects the late time hydrodynamic behavior of quantum systems to be largely classical \cite{hohenberg1977theory}. Much less is known when such systems are driven far from equilibrium, especially in the presence of  \textit{macroscopic} manifestations of the uncertainty intrinsic to quantum mechanics at equilibrium. A concrete setting in this context, and our focus in this article, are regimes of hydrodynamics where the many-particle quantum dynamics exhibits collective phenomena similar to that of turbulent fluids.   

\begin{figure}
    \centering
\includegraphics[width=0.8\linewidth]{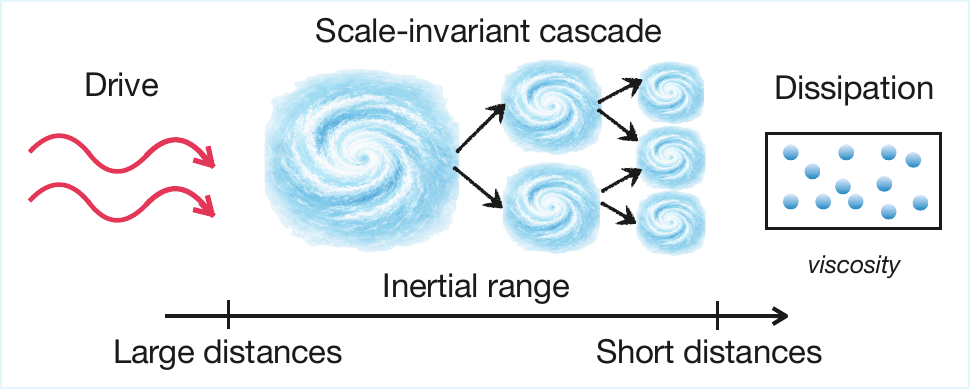}
    \caption{A cartoon of turbulence in a classical fluid in the fully-developed regime. An external drive (a force $\bm{f}$) pumps in energy at large distances (such as the fluid system size $L$) in a fluid. This leads to a \textit{scale-invariant} cascade over an 'inertial range' until the energy is dissipated at short distances by viscous forces. The whirls in the inertial range denote eddies, which dump energy into smaller ones \textit{without} dissipation throughout the inertial range.}
    \label{fig:classical_turbulence_sketch}
\end{figure}

\begin{figure}
    \centering
\includegraphics[width=0.95\linewidth]{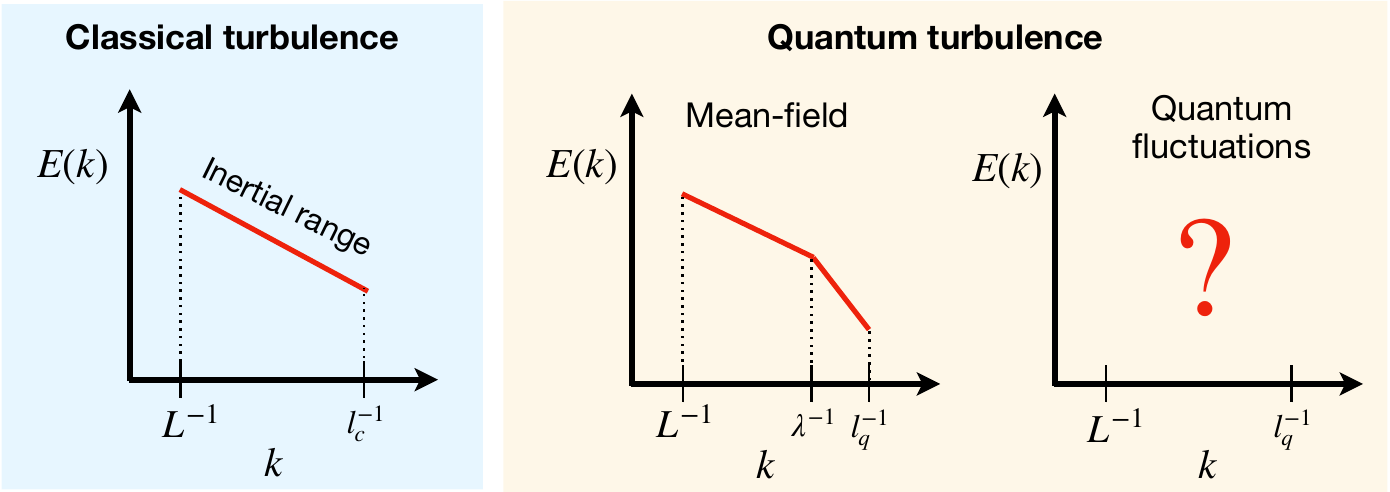}
    \caption{Sketches of the one-dimensional energy spectrum in classical and quantum fully-developed turbulence, as a function of wavenumber $k$ in the inertial range on a double logarithmic scale. The question mark indicates our lack of understanding of quantum turbulence in fluids with strong quantum fluctuations, indicating a new frontier for the many-body quantum dynamics. $L$ is the large-scale system size of the fluid, while $l_c$ or $l_q$ is a small length-scale related to dissipation. $\lambda$ denotes the mean inter-vortex spacing, which often sets the knee for the energy spectrum in quantum turbulence in three dimensions in the mean-field approximation. Typically $l_c$ is set by viscosity; in Kolmogorov's theory, $l_c\propto \nu^{3/4}\epsilon^{-1/4}$, where $\nu$ is the dynamic viscosity and $\epsilon$ is the rate of energy transfer. In quantum turbulence, at mean-field level, $l_q\sim \xi$, where $\xi$ is the vortex core size (healing length).}
    \label{fig:sketch}
\end{figure}

Classical turbulence is a phenomenon characterized by chaotic dynamics of eddies, coupled with an efficient transfer of energy from large-scale injection to small-scale dissipation \cite{Frisch:book,Verma:book:ET}. Traditionally described by the Navier-Stokes equations, these flows rely on viscosity to balance the energy input from driving by dissipating kinetic energy as heat over time. For concreteness, consider a driven classical fluid—in such systems, at high flow velocities, an isotropic homogeneous steady-state with universal dynamics may be obtained. In the so-called \textit{fully-developed} turbulence, one observes a power-law scaling in equal time two-point spatial correlations functions, say, of the fluid velocity. This is called a cascade and is interpreted as a rapid transfer of a physical quantity across space. In this case, e.g., the fluid receives energy from the drive at long wavelengths, which cascades from the infrared (IR) to the ultraviolet (UV) (smallest wavelengths) in a scale-invariant manner (see Fig.~\ref{fig:classical_turbulence_sketch} for a cartoon picture of the Richardson's interpretation of a cascade~\cite{richardson1922weather} and Fig.~\ref{fig:sketch}(a) for an illustration of the associated power law). At the ultraviolet, energy is dissipated typically by viscous damping. For a three-dimensional classical fluid, steady-state turbulence yields the following celebrated power law,
\begin{equation}\label{kolmogorov_scaling}
    \langle v_kv_{-k}\rangle\sim k^{-5/3},
\end{equation}
where $v_k$ is the velocity amplitude at wavenumber $k$, and $\langle.\rangle$ denotes a time average\footnote{A more precise definition is as follows. $E(k)$ is defined through the equal-time two-point correlation function of the velocity $C_{\alpha\beta}(\bm{r})=\langle v_\alpha(\bm{r},t)v_\beta(\bm{0},t)\rangle$, with the definition, $\sum_\alpha C_{\alpha\alpha}(\bm{0})=\int_0^\infty \textrm{d} k\: E(k)$, where $\alpha,\beta$ denotes the coordinate axes.}. A theoretical prediction of this power law (and the universal coefficient) from a model of the classical fluid has proven to be a notoriously difficult task.

The question of turbulence in quantum fluids arose through the demonstration of superfluidity in Helium-4 \cite{allen1938flow, kapitza1938viscosity}, and subsequently in cold-atom condensates \cite{anderson1995observation, davis1995bose}. In superfluids, quantum mechanics constrains the hydrodynamics by a quantization of the circulation of the velocity in multiples of the ratio of Planck's constant to the mass of the boson. 
Near zero temperature, dissipation occurs through radiation due to phonons at the microscopic scale. Energy is transported to this scale, in three dimensions, via helical oscillations of vortex lines---the topological defects of the order parameter---known as Kelvin waves \cite{kozik2004kelvin}.
Over the past couple of decades, theoretical, numerical, and experimental efforts have uncovered (to a great extent) the physics of such quantum turbulence \cite{Barenghi:book:QT}. In three dimensions, two distinct cascade regimes are typically obtained, interjected by a length-scale corresponding to an average inter-vortex spacing (see Fig.~\ref{fig:sketch}(b) for an illustrative example). These studies, however, have been largely limited to the regime of \textit{small quantum fluctuations}, where mean-field approximations are sufficient. Indeed, a typical tool for analysis is a classical hydrodynamical equation known as the Gross-Pitaevskii equation \cite{gross1961structure, pitaevskii1961vortex}, whose evolution can be readily computed in a classical computer \cite{Barenghi:book:QT}.

We can make the notion of `small quantum fluctuations' more precise. The superfluid is a quantum phase of matter and a ground state of a many-body quantum Hamiltonian. Nevertheless, deep within a superfluid, the fluctuations are largely classical, since the fluctuations in the number of bosons locally are small compared to the average boson density. Thus, in this phase, fluctuations induced by Heisenberg's equation of motion are necessarily small. Therefore, it seems natural to suggest that quantum fluctuations play a negligible role in the large scale hydrodynamic behavior of typical superfluids. 

By contrast, there exist quantum fluids where quantum fluctuations are strong, and play a crucial role in determining macroscopic physical observables. An example of a fluid where strong quantum fluctuations are realized is in a system of bosons on a lattice at zero temperature, near a phase transition marking the loss of superfluidity \cite{sachdev1999quantum, fisher1989boson}. Strictly speaking, a lattice breaks the conservation of momentum; however, at length-scales much bigger than the lattice spacing, a hydrodynamic description should suffice. Near the phase transition, a mean-field treatment breaks down and one has to incorporate quantum fluctuation effects in order to obtain correlation functions, both in and out of equilibrium. It thus seems likely that the hydrodynamic behavior of such a quantum critical fluid, including its turbulent dynamics, can show macroscopic signatures of many-body quantum effects. This would be a `many-body regime' for quantum turbulence, as stated in the title of this article. We provide more examples of candidate fluids with strong quantum interactions in Sec.~\ref{many-body} and Sec.~\ref{discussion}. While a hydrodynamic formulation of the dynamics of any classical or quantum many-body system is typically in terms of classical fields, for systems with large quantum fluctuations, identifying the useful classical degrees of freedom may itself prove to be a non-trivial task. We elaborate more on this point in Sec.~3\ref{lattice_model} and explain why such a description may require a quantum many-body perspective.

A primary goal of our article is to suggest that turbulence in fluids with strong quantum fluctuations can realize a new frontier for quantum turbulence. Our discussion in this article will be two-fold: (i) a concise review of the phenomenology of quantum turbulence (as studied so far), and (ii) an identification of quantum fluids with strong quantum fluctuations, with suggestions for calculations that may uncover quantum many-body effects in their turbulent hydrodynamics. Since readers familiar with quantum many-body theory may not be acquainted with the phenomenology of turbulence and {\it vice-versa}, we treat this article as a means to synthesize the relevant know-how, setting the stage for future work. (We do not review turbulence in classical fluids, but the reader may refer to standard books such as Ref.~\cite{Frisch:book}.) With rapid progress in cold-atom experiments and quantum computing platforms, we find it very plausible that our set of questions can also be answered in present-day experiments.

The organization of this paper is as follows. 
In Sec.~\ref{sec:quantum_fluids}, we review turbulence in superfluids, focusing on the mean-field model known as the Gross-Pitaevskii equation. We discuss the relation of hydrodynamics of this fluid to that of a compressible fluid with zero viscosity, along with consequences for its turbulent dynamics. The informed reader may skip this section and directly read the next section, Sec.~\ref{many-body}. Here, we consider the question of many-body quantum effects in fluid turbulence with strong quantum fluctuations, and provide concrete examples where this may be studied. 
In Sec.~\ref{discussion}, we synthesize our discussion and suggest new directions, of interest to modern developments in non-equilibrium quantum dynamics.

\section{Turbulence in quantum fluids: mean field}\label{sec:quantum_fluids}

There has been a long history of the study of hydrodynamics in quantum fluids.  A quantum fluid, broadly defined, is a system that exhibits quantum effects at macroscopic length and time scales.  For instance, at low temperatures, there is the tendency of classical systems to solidify; by contrast, in Helium-4 (below a condensation temperature $T_c\approx 2$ K, and at ambient pressure), the zero-point motion due to Heisenberg's uncertainty principle prevents crystallization and the system instead forms a Bose-Einstein condensate \cite{leggett2006quantum}. The superfluid phase is a boson condensate and exhibits striking phenomena such as dissipation-less flow and a circulation that is quantized in multiples of the ratio of Planck's constant to the mass of the boson \cite{pitaevskii2016bose, svistunov2015superfluid}.

While the thermodynamics of quantum fluids requires quantum mechanics, it is by no means obvious that the hydrodynamics that governs the large scale behavior of such systems is intrinsically quantum mechanical. In what follows, we shall concisely review to what extent the laws of quantum mechanics enter the macroscopic turbulent hydrodynamics of superfluids, even within a mean-field approximation. We restrict our discussion to three spatial dimensions. Here, we find that quantum effects \textit{only} enter the turbulent hydrodynamics at scales smaller than a mean inter-vortex spacing $\lambda$, and at temperatures much smaller than $T_c$. 

We shall discuss two dynamical regimes of interest, which are classified as per their temperature scale \cite{barenghi2014introduction}\footnote{A definition of temperature for non-equilibrium classical and quantum systems, especially when not in a steady-state is a non-trivial task (see also the discussion in Sec.~\ref{discussion}). Here, we use the notion of temperature in a limited sense, only to denote the fraction of bosons participating in the condensate. When $T\rightarrow 0$, the superfluid fraction is $1$, while when $T=T_c$, the superfluid fraction $\rightarrow 0$. Formalizing the notion of temperature of out-of-equilibrium (e.g. steady-state turbulent) hydrodynamics from a quantum statistical mechanics point-of-view may be an interesting unresolved question.}. (i) Superfluid turbulence at zero temperature $(T=0)$. This is to be regarded as an isolated system (such as the Euler fluid), and zero temperature indicates that nearly all of the bosons participate in the condensate [the normal (viscous) fluid is absent]. Turbulent cascades can be seen in such fluids even within Hamiltonian evolution, starting from an appropriate initial state and boundary condition. (ii) Next, we briefly consider superfluid turbulence at finite temperatures. The temperatures considered will always be smaller than the condensation temperature ($T_c$), so that there is a finite superfluid fraction. When $0<T\ll T_c$, a small fraction of the bosons is lost from the condensate, leading to dissipation in the superfluid. Such a superfluid can be driven to obtain a steady-state fully-developed turbulent regime here, akin to the Navier-Stokes equation. At even larger temperatures, the dissipated phonons can themselves thermalize, and form a normal fluid, which can undergo its own turbulent dynamics as well. When we discuss the scope for many-body turbulence in quantum fluids in Sec.~\ref{many-body}, we discuss counterparts to both of the dynamical regimes.

\begin{figure}
    \centering
    \includegraphics[width=0.95\linewidth]{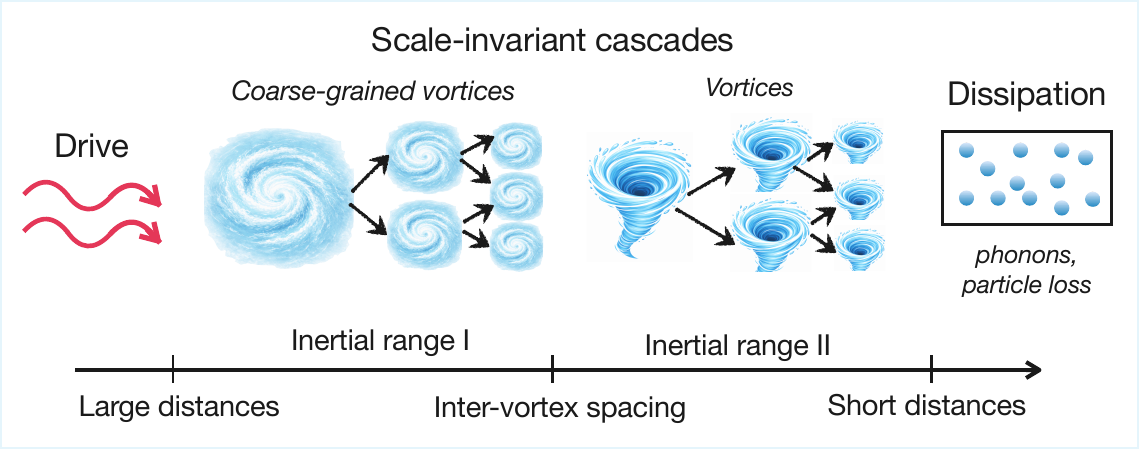}
    \caption{A schematic of fully-developed turbulence in a three-dimensional superfluid, particularly in the regime $0<T\ll T_c$. Here, one can have two inertial ranges, separated by a length-scale related to the inter-vortex spacing $(\lambda)$. Inertial range I (for length-scales between system-size $L$ and $\lambda$) exhibits a Kolmogorov-like spectrum with exponent $-5/3$, and may be interpreted as a Richardson cascade of coarse-grained vortices (eddies)~\cite{richardson1922weather}. Inertial range II [for length-scales between $\lambda$ and vortex core size (healing-length) $\xi$] exhibits a spectrum obtained from the dominant process in vortex dynamics, such as a Kelvin wave cascade. Dissipation at length-scales smaller than $\xi$ may occur through phonons released during vortex dynamics, and/or particle loss from the experimental setup.}
    \label{fig:quantum_sketch}
\end{figure}

\begin{figure}
    \centering
    \includegraphics[width=0.85\linewidth]{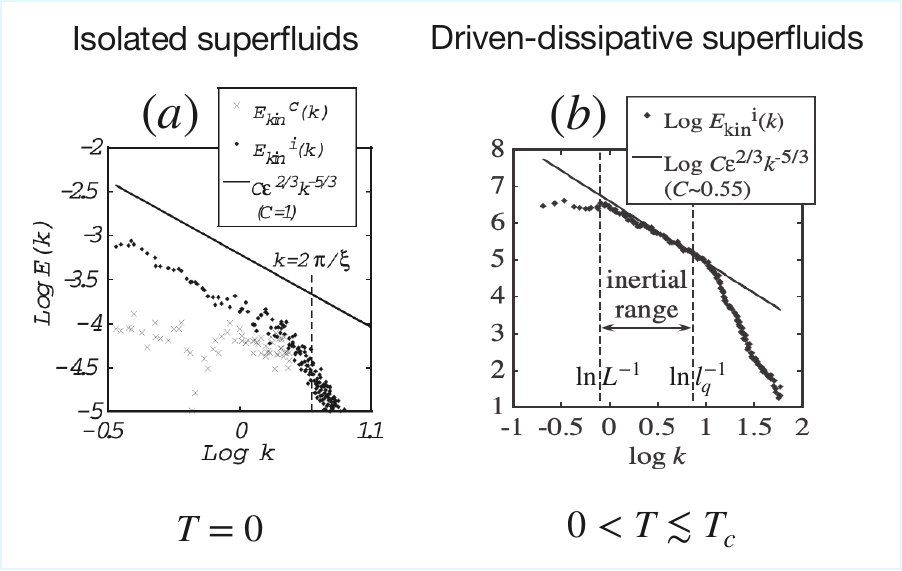}
    \caption{\textbf{(a)} [Figure reproduced with permission from Ref.~\cite{kobayashi2005kolmogorov}] Kolmogorov-like turbulence in a three-dimensional GP fluid without driving. The initial state chosen was of uniform density and random phase. The incompressible part of the spectrum $E_\textrm{kin}^i(k)$ (defined using a density-weighted incompressible velocity field [see Ref.~\cite{kobayashi2005kolmogorov} for details] exhibits the $-5/3$ law over an inertial range I at short times (the data plotted is at $\sim \mathcal{O}(1)$ time units, defined by the ratio of the healing length to the speed of sound). (\textbf{b}) [Figure reproduced with permission from Ref.~\cite{kobayashi2005kolmogorov2}] Fully-developed turbulence exhibiting the Kolmogorov cascade (inertial range I) in direct numerical simulations of a driven-dissipative GP equation. The driving potential is a random moving potential with two-point Gaussian correlations. The inertial range is defined between a large-scale wavenumber $L^{-1}$ corresponding to the random potential, and a small-scale dissipation scale $l_q^{-1}$ corresponding to the phenomenological dissipation.}
    \label{fig:quantum_plots}
\end{figure}

The Hamiltonian of a system of bosons whose ground state can admit a superfluid phase is given by (in second-quantized notation),
\begin{equation}\label{second_quantized_bosons}
\hat{H}
=
\int_r\;
\hat{\psi}^\dagger(\mathbf r)\left(-\frac{\hbar^2}{2m}\nabla^2\right)\hat{\psi}(\mathbf r)
\;+\;
\frac{1}{2}\int_{r,r'}
\hat{\psi}^\dagger(\mathbf r)\hat{\psi}^\dagger(\mathbf r')\,
V(\mathbf r-\mathbf r')\,
\hat{\psi}(\mathbf r')\hat{\psi}(\mathbf r),
\end{equation}
where $m$ denotes the mass of the boson, $V$ denotes the interaction between the bosons and $\int_r=\int\textrm{d}^3\bm{r}$. $\hat{\psi}$ and $\hat{\psi}^\dagger$ are the boson annihilation and creation operators at position $\bm{r}$. We shall set $m=1$ ahead for a simplified presentation. The second-quantized boson field operators obey the equal time commutation relation,
\begin{equation}\label{commutation_relation}
\left[ \hat{\psi}(\bm{r}, t), \hat{\psi}^{\dagger}(\bm{r}',t) \right] = \im  \hbar \delta^3(\bm{r}-\bm{r}').
\end{equation}
Studying this problem in its full scope is unfeasible given the exponential size of the Hilbert space. A useful approximation can be made when quantum fluctuations can be ignored, often known as mean-field theory. Deep within the superfluid phase, the fluctuations in boson number locally is negligibly small compared to the average condensate density and thus, such mean-field treatments ought to capture the dynamics adequately.

A convenient mean-field theory of the system of bosons in Eq.~\ref{second_quantized_bosons}, where $V$ denotes contact repulsive interactions of strength $g$, is known as the Gross-Pitaevskii (GP) equation. It may be obtained within the Hartree approximation of the many-body ground state \cite{pitaevskii2016bose}. The resulting equation is a nonlinear Schr\"{o}dinger equation for the complex order parameter $\psi({\bf r},t)$ \cite{gross1961structure, pitaevskii1961vortex}, where the nonlinearity comes from boson interactions. The complex field $\psi(\bm{r},t)$ may be interpreted as the expectation value of the boson annihilation operator in a product state. The GP equation is given by,
\begin{equation}\label{gpe}
    \textrm{i} \hbar \del_t\psi = -\frac{1}{2}\nabla^2\psi+g|\psi|^2\psi.
\end{equation}

The GP model is particularly suitable to describe dilute cold-atom condensates and is reasonable description of Helium-4. The contact repulsion approximates the rapidly decaying van der Waals interaction between the neutral atoms. However, the low-energy excitation spectrum of Helium-4 has a minimum at finite momentum (known as the roton), as observed in experiments, which the GP equation fails to capture \cite{pitaevskii2016bose}.

Regardless, the GP equation admits vortex solutions, which in three dimensions are extended one-dimensional objects. At a vortex, the density $\rho(\bm{r},t)$ falls to zero, over a coherence length (also known as the healing length) $\xi=\sqrt{\hbar^2/2g\rho_0}$, where $\rho_0$ is the average density. Around a vortex, the phase $\theta(\bm{r},t)$ of $\psi$ winds non-trivially. Specifically, a vortex has a quantized circulation, that is, 
\begin{equation}
        \oint_\gamma \frac{\hbar}{m}\bm{\nabla}\theta\cdot \textrm{d} \bm{l}= \frac{h}{m}q, \;\;\; q\in \mathbb{Z},
\end{equation}
where $\gamma$ denotes a contour around the vortex, $\textrm{d}\bm{l}$ is the infinitesimal line segment and we have temporarily reintroduced $m$ in this equation. The integer $q$ is the `charge' of the vortex, with the $q=1$ vortices being the most energetically favorable. This constraint arising from quantum mechanics leads to a non-trivial consequence in the turbulent cascade spectrum.

At zero temperature, nearly all of the bosons participate in the condensate, yielding a superfluid with zero viscosity. This is similar to the Euler fluid, a comparison that can be made precise as explained ahead. In the GP model, the low-energy excitations are acoustic phonons with a linear dispersion $[\omega(k)]$ (corresponding to superfluid density variations) at small momentum $k$: $\omega(k)\sim c_sk$, where the speed of sound $c_s=\sqrt{g\rho_0}$~\cite{sachdev2023quantum}. These phonons help provide an emergent dissipation to the incompressible part of the kinetic energy, even at zero temperature (note that the full system is closed, and hence conserves energy). This energy to the phonons is typically transported through vortex dynamics, such as reconnection and helical oscillations of the vortex lines, known as Kelvin waves \cite{barenghi2014introduction}. The GP model is able to account for such non-trivial vortex dynamics. 

Particularly in numerical simulations, the following hierarchy of length-scales is often realized,
\begin{equation}
    L\gg \lambda\gg \xi,
\end{equation}
where $L$ is the system size, and $\lambda$ and $\xi$ are the mean inter-vortex spacing and healing length respectively. In such a case, two inertial ranges [inertial range I and II of Fig.~\ref{fig:sketch}(b) and Fig.~\ref{fig:quantum_sketch}]---wave number ranges over a power law scaling is observed---emerge, with different turbulent power laws. Exactly at zero temperature, the lack of an explicit dissipative mechanism implies that the dynamics in this regime is unforced, since forcing the fluid will lead to an unbounded increase in energy. In this case, one can prepare the superfluid in a non-equilibrium state, and evolve the system to obtain a turbulent cascade at early times. At long times, the cascade in the inertial range typically crossovers to an equilibrated spectrum (similar to the dynamics of an Euler fluid). This regime may be interpreted as the study of turbulent dynamics after a quench, as often studied in quantum many-body systems \cite{polkovnikov2011colloquium}. 

Numerical simulations of the GP equation reveal that in inertial range I, the kinetic energy spectrum is typically consistent with Kolmogorov's law ($k^{-5/3}$) in three dimensions \cite{nore1997decaying, nore1997kolmogorov} (see e.g. Fig.~\ref{fig:quantum_plots}(a)). In a sense, one can interpret this as if at these length-scales the vortex dynamics is averaged over, and the resulting eddies undergo a classical Richardson cascade (see Fig.~\ref{fig:quantum_sketch}).

For length-scales smaller than $\lambda$, Feynman hypothesized that the dynamics of the vortex lines, often involved in a complicated tangle, can lead to a power law different from $-5/3$ \cite{feynman1955chapter}. Indeed, the cascade in this regime is often dominated by wave-like interactions between helical oscillations of the vortex lines, known as a Kelvin wave cascade. Remarkably, this limit admits an analytical calculation of the cascade power law, using the kinetic theory of (weak) wave-turbulence \cite{kozik2004kelvin, l2010spectrum, nazarenko2011wave}. The power law computed by L'vov and Nazarenko (amusingly with the same exponent $-5/3$) has been verified in Gross-Pitaevskii numerics [see Ref.~\cite{krstulovic2012kelvin}].

We now return to the simpler fact of the emergent classical behavior (Kolmogorov law) at large length-scales ($>\lambda$). As it turns out, the GP equation can be rewritten as a hydrodynamic equation using the so-called Madelung transformation \cite{madelung1927quantum}, that writes the order parameter $\psi(\bm{r},t)=\sqrt{\rho(\bm{r},t)}\textrm{exp}({\textrm{i}\theta(\bm{r},t)})$ in terms of density and phase fields. Then, one can derive the following two equations, \cite{pitaevskii2016bose},
\bea
&\partial_t \rho  +\bm{\nabla}\cdot (\rho\bm{v})=0, \label{quantum_continuity}\\
&\rho(\del_t+\bm{v}\cdot\bm{\nabla})\bm{v}=-\rho\bm{\nabla}{Q(\rho)}-\bm{\nabla}\left[\frac{g}{2}\rho^2\right]\;\;\; \textrm{where}\;\;\;  Q(\rho)=-\frac{\hbar^2}{2m}\frac{\nabla^2\sqrt{\rho}}{\sqrt{\rho}},
    \label{eq:GP_u}
\eea
which hold away from vortices, and we have defined a fluid velocity $\bm{v}=\hbar\bm{\nabla}\theta$.  Note that this implies that away from the vortices, the fluid velocity in the superfluid is irrotational. Eq.~\ref{quantum_continuity} is just the continuity equation corresponding to particle number conservation of the superfluid. Eq.~\ref{eq:GP_u} is a compressible Euler equation, with a quantum-mechanical correction $\propto \bm{\nabla}Q(\rho)$. $Q(\rho)$ is a quantum mechanical energy (sometimes also called the quantum potential \cite{ballentine2014quantum}). Upon rewriting the energy of the condensate in terms of the hydrodynamic variables, one may see that the role of this potential is to suppress density variations, such as those that occur at a vortex. 
 
One may compare the relative importance of the force coming from the quantum mechanical potential to that from boson-boson interactions by considering the following ratio, which may be called a Thomas-Fermi parameter (TF),
\begin{equation}\label{ratio}
\textrm{TF}\approx \frac{|\rho_0\bm{\nabla}Q(\rho_0)|}{\bm{\nabla}[g\rho_0^2]}\approx \frac{\hbar^2}{2 g \rho_0} \frac{1}{l^2}=\left(\frac{\xi}{l}\right)^2,
\end{equation}
where we have assumed that the average density $\rho_0$ varies over a length-scale $l$. The so-called Thomas-Fermi limit, at large $g$, is when $\textrm{TF}\rightarrow 0$ \cite{pethick2008bose}. Therefore, at long wavelengths, such as when $l\sim O(10)\xi$ (and certainly at length-scales bigger than the inter-vortex spacing), the interaction energy dominates over the quantum potential energy and we can safely ignore this term. We may now directly compare the resulting Eq.~\ref{eq:GP_u} to that of a compressible classical inviscid fluid, 
    \begin{equation}
\rho(\del_t+\bm{v}\cdot\bm{\nabla})\bm{v}=-\bm{\nabla}p(\rho,s),
\end{equation}
whose pressure $p(\rho,s)$ depends on the density $\rho$ and the entropy density $s$ by an equation of state. For a classical fluid undergoing adiabatic dynamics, the equation of state is $p(\rho,s)\propto \rho^\gamma$, where $\gamma$ is the ratio of the specific heats at constant pressure and volume. Comparing with the interaction pressure $(g/2)\rho^2$, it is clear that in the superfluid dynamics at $T=0$, the role of temperature is played by the interaction strength $g$, and exhibits hydrodynamics similar to an ideal gas with $\gamma=2$. We shall return to this comparison with the Euler fluid briefly at the end of Sec.~4\ref{lattice_model}, when we discuss the role of quantum fluctuations in turbulence.

Upon heating the fluid above zero temperature, but still at temperatures much smaller than the condensation temperature, additional sources of dissipation arise from thermal excitations (and weak contact with a thermal reservoir). Given an explicit sink for dissipation, it is now possible to drive the fluid and achieve a steady-state in the sense of fully-developed turbulence. This is a regime that is accessible in several cold-atom condensates \cite{navon2016emergence, navon2019synthetic}. 

The phenomenology of turbulence observed in this regime is similar to that at $T=0$, except that cascade power laws can now be obtained in the steady-state \cite{kobayashi2005kolmogorov2}. For temperatures much smaller than $T_c$, often Gross-Pitaevskii-like description continues to be useful, albeit with a phenomenological dissipative term (see Ref.~\cite{bradley2012energy} and references within). In Ref.~\cite{kobayashi2005kolmogorov2}, a turbulent steady-state was realized in a driven GP equation with a dissipation that acts at short-distance wavelengths $\sim\xi$, and a Kolmogorov spectrum was indeed obtained (see Fig.~\ref{fig:quantum_plots}(b)).

At even higher temperatures (typically $T\gtrsim T_c/2$), the turbulent dynamics of the normal fluid need to be considered as well. This is often the regime that is accessible in Helium-4 experiments \cite{skrbek2021phenomenology}. A convenient description of the dynamics here is using Landau's two fluid model \cite{landau2018theory}, describing the coupled super- and normal fluids. Like at small but finite temperatures, a source of dissipation is the mutual friction between the fluids. At such high temperatures, quantum effects are typically washed out and one obtains a spectrum consistent with Kolmogorov's law (see e.g. Ref.~\cite{boue2015energy} for a detailed analysis of this regime).

\section{Turbulence in quantum fields: beyond mean-field}\label{many-body}

In the sections above, we briefly reviewed the problem of turbulent hydrodynamics in superfluids, ignoring  quantum fluctuations, which are small deep within the superfluid phase. However, quantum fluctuations play an important role in determining the thermodynamics and dynamics when the system is tuned towards a continuous zero temperature phase transition.  In this regime, boson density fluctuations can be comparable in magnitude to the mean boson density, and at criticality, quantum fluctuations stemming from Heisenberg's equations of motion cannot be neglected. Consequently, one must go beyond mean-field theory in determining the dynamics.  As we discuss below, many-body effects become crucial in determining the hydrodynamic behavior near criticality. Our goal in this section is to detail circumstances under which turbulent hydrodynamics of a quantum fluid may show quantum {\it many-body} behavior. We shall discuss this idea at a broad level, leaving scope for a detailed investigation in future work.  

We should first clarify what we mean by a `quantum many-body' description of the system.  In analyzing the Gross-Pitaevskii equation, one interprets the variable $\psi$ as the wave-function governed by a non-linear Schrodinger equation. While it accounts for many-particle effects using the nonlinear term arising from interactions, ultimately, this equation governs the dynamics of the wave-function $\psi$ of a {\it single} particle. Information of many particles may be obtained by following one of two routes. In the first route, we can promote the complex field $\psi(\bm{r},t)$ to a {\it field operator} $\hat{\psi}(\bm{r},t)$ (thus denoted with a hat) that lives in a many-body Hilbert space and destroys a boson at position $\bm{r}$ at time $t$. This is the sense in which the second-quantized formulation of the Hamiltonian in Eq.~\ref{second_quantized_bosons} is defined.  The field $\hat{\psi}$ operator couples a Hilbert space of $N$ bosons to that of $N-1$ bosons.  Similarly, one can promote the complex conjugate $\psi^*(\bm{r},t)$ to a field operator $\hat{\psi}^\dagger(\bm{r},t)$ that creates a boson at $(\bm{r},t)$.  The creation operator mixes the $N$ body Hilbert space and the $N+1$ body Hilbert space.  By considering a theory in many such Hilbert spaces with differing number of particles (known as a ``Fock space"), we arrive at a truly many-body description of the Bose fluid. 

A second route to the same endpoint involves thinking of $\psi$ not as a wavefunction, but rather as a complex scalar {\it field}, that lives everywhere in spacetime.  The GP equation can be thought of as the equation of motion coming from a classical field theory.  To arrive at a quantum many-body version of the GP equation, we promote the classical field theory to a quantum field theory by imposing the Heisenberg uncertainty principle on the field $\psi, \psi^*$.  These are the commutation relations written in Eq.~\ref{commutation_relation}. Then, the theory to study is the full quantum action of bosons whose equation of motion is the GP equation. This more general theory involves path integration over all complex valued scalar field configurations, not just those that belong to the mean-field saddle point.   

The most important  quantum fluctuations are in the number of bosons at each position.  
To realize a superfluid with strong quantum fluctuations, it is helpful to consider 
 a lattice formulation in which bosons are tunneling across neighboring bonds in the presence of repulsive, short-ranged interactions (for example, see the discussion of the Bose-Hubbard model in Sec.~4\ref{lattice_model}). In this case, quantum fluctuations lead to a non-zero root-mean-squared fluctuation in the boson number at each lattice site: $\delta n \equiv \sqrt{\langle \hat{n}^2 \rangle}$.  If this quantity is small compared to the average boson density, namely if $\delta n \ll \langle n \rangle$, as is the case deep within a superfluid, then quantum fluctuation effects can be neglected and the GP equation does an adequate job in capturing hydrodynamic behavior of the superfluid.  By contrast, if $\delta n \gtrsim \langle \hat{n} \rangle$, then quantum fluctuations cannot be ignored and play a crucial role in the ensuing hydrodynamic behavior.  Such a system is poorly described by the GP equation.  
 
The most natural place to find such behavior would be near a zero temperature phase transition marking the loss of superfluidity.  By definition, at this transition the expectation value $\langle \hat{\psi} \rangle = 0$ and if the transition is continuous, fluctuations in $\hat{\psi}$ occur over all length and time scales.  Such effects are invisible within the GP equation and a lattice formulation is crucial, because the phase with $\langle \hat{\psi} \rangle = 0$ is a Mott insulator, which requires a underlying lattice. The finite temperature behavior of such a system, when a thermal length scale $\lambda_T \sim h/k_B T$ ($k_B$ is Boltzmann's constant) is small compared to the boson correlation length, presents a novel hydrodynamic regime that has not been extensively explored in the context of  quantum turbulence.  If the system in this regime is stirred out of equilibrium, any resulting turbulent hydrodynamics would be, by definition, a many-body quantum turbulent regime, one that would be poorly captured by the GP equation.

To be concrete, let us consider the example of a system of bosons in two spatial dimensions tuned to criticality, say, the Wilson-Fisher fixed point \cite{Wilson:PR1974}. We emphasize that here we refer to the \textit{zero temperature} quantum critical point described by the $O(2)$ Wilson-Fisher theory, describing the onset of superfluidity. This fixed point is described by a relativistic quantum field theory of a complex scalar. The turbulent hydrodynamics of such a fluid can have different (emergent) conserved quantities as compared to its single-particle counterpart, and this may lead to power laws that are distinct from those of its single-particle quantum counterpart, and certainly of the classical fluid. At equilibrium, in the case of the Wilson-Fisher fixed point, fluctuations in the order parameter lead to spontaneous production and annihilation of vortices from the vacuum. This is otherwise absent in the superfluid phase, where creation of vortices cost a finite amount of energy. Upon driving such a quantum critical fluid, it is conceivable that the non-conservation of vortex number may lead to a distinct scaling structure. As we have discussed before, in Gross-Pitaevskii turbulence, two regimes of power laws are typically observed---at length scales bigger than a mean inter-vortex spacing, the scaling is classical (inertial range I), while below this length scale vortex dynamics governs the scaling (inertial range II), leading to a scaling law absent in a classical Navier-Stokes fluid. In the turbulent dynamics of the Wilson-Fisher fluid, it is possible that the lack of vortex conservation may imply no length scale associated to an average vortex density, hence one may obtain a single scaling regime. Owing to an emergent Lorentz invariance, this power law may be distinct from both superfluid and classical fluid turbulence. 

There is another way to think about the turbulence of a quantum many-body fluid that suggests a new scaling law. For equilibrium quantum field theories, the equation of motion of the degrees of freedom in a phase is obtained by evaluating the saddle-point of the corresponding path integral. As one approaches a phase transition, quantum fluctuations renormalize classical scaling dimensions of operators, leading to critical scaling laws. By analogy, it is possible that quantum fluctuations that are dominant near the Wilson-Fisher fixed point may renormalize scaling of operators, whose coarse-grained expectation values enter a hydrodynamic equation and lead to new turbulent scaling laws. More precisely, quantum fluctuations of the field $\psi(\bm{r},t)$ may lead to renormalization of the dimensions of the velocity field (which deep within the superfluid phase is given by the gradient of the phase of $\psi$), and thus affect any advective term that leads to cascades. Of course, the turbulent steady-state in such a strongly-interacting fluid will likely have a critical scaling different from the \textit{equilibrium} critical scaling due to renormalization from interactions. However, the fact that the unforced equilibrium state undergoes strong quantum fluctuations suggests that the power law obtained in steady-state turbulence of the driven fluid may be distinct from that of the weakly-interacting superfluid. 

In Kolmogorov's theory, in addition to a universal power law, there is a dimensionless prefactor (Kolmogorov constant) that is also found to be universal in steady-state dynamics~\cite{sreenivasan1995universality}. Another interesting question is to consider the fate of this constant in quantum many-body turbulence. Since it must be dimensionless, it is possible that now the constant is \textit{tunable}---in fact, it may depend on the ratio of the inverse mass of the boson to the strength of its repulsive interactions, or, in other words, the fine structure constant of the system.

We now provide two concrete examples: (i) in Sec.~4\ref{lattice_model}, we describe a model of bosons on a lattice, known as the Bose-Hubbard model, that shows a quantum phase transition at zero temperature. We primarily discuss this phase transition in two dimensions. (ii) We also consider a model of bosons in one dimension. Quantum fluctuations tend to be stronger in low dimensional systems; consequently, we do not expect the long-distance behavior of a one-dimensional Bose fluid to be well described by the GP model. In particular, the lack of long-range order in one dimension (as per the Mermin-Wagner theorem \cite{mermin1966absence}) implies that the one-dimensional fluid has power law correlations. We outline our interest in this setting in Sec.~4\ref{one_d_problem}.

\subsection{Turbulence in the vicinity of a quantum phase transition}\label{lattice_model}

\begin{figure}
    \centering
\includegraphics[width=1\linewidth]{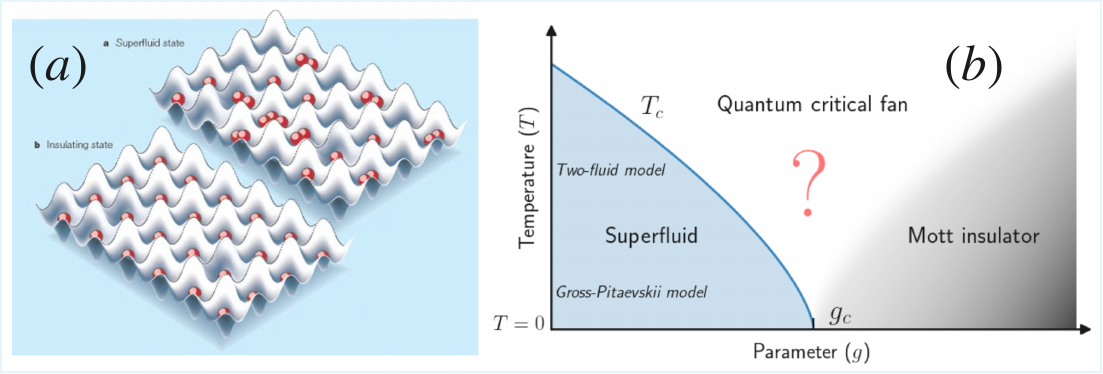}
    \caption{\textbf{(a)} [Figure reproduced with permission from Ref.~\cite{stoof2002breaking}]. A cartoon picture of a superfluid and Mott insulator in two dimensions. In the superfluid, the bosons (denoted by red spheres) achieve phase coherence, with density fluctuations small compared to the average density. In the insulator, the ground-state is a density eigenstate, with a fixed number of bosons (one, in this case) per lattice site. The quantum critical fluid sits between these two phases, where fluctuations exist at all length-scales. \textbf{(b)} Finite temperature phase diagram of the two-dimensional Bose-Hubbard model at the particle-hole symmetric point, which belongs to the $O(2)$ Wilson-Fisher universality class. The $x-$axis denotes a parameter such as $g$ proportional to the interaction strength in Eq.~\ref{bose-hubbard}. The $y-$axis denotes temperature. The quantum phase transition at $T=0$ occurs at a critical value $g=g_c$. The blue line indicates a continuous phase transition between the superfluid and the normal fluid at a temperature $T_c$. Between the superfluid (blue) and Mott insulator (grey), there exists a quantum critical fan, which is a quantum fluid with strong quantum fluctuations. The position of the labels 'two-fluid model' and 'Gross-Pitaevskii model' schematically indicate their likely regime of applicability.}
    \label{fig:many-body}
\end{figure}

A convenient way to study the physics of superfluids is to restrict the order parameter $\psi$ to be defined on a discrete spatial grid (i.e. a lattice).  The price we pay in doing this, is that we lose the strict sense of momentum conservation, since continuous translation symmetry is broken.  We thus expect that any hydrodynamic behavior in such systems would only occur at wavevectors $k \ll 1/a$, where $a$ is the lattice spacing. One can also restrict to dilute densities to suppress lattice effects, which occurs when the mean-free path of the bosons is large compared to the lattice spacing. 

Including a lattice, however, gives us the ability to describe phases that would result in the strong coupling limit where interactions dominate over the kinetic energy.  Let $\psi_i$ be the amplitude of the boson order parameter on the lattice site labeled by subscript $i$, and let the complex conjugate be denoted by $\psi^{\dagger}_i$.  For the time being, these represent complex numbers defined on lattice sites, but later on, they will represent quantum Bose fields. The analog of the GP equation (in Eq.~\ref{gpe}) for such a system is obtained by exchanging the continuum Laplacian for a lattice counterpart, which is merely the lattice difference, 
\begin{equation}\label{lattice_gpe}
    \textrm{i} \hbar \partial_t \psi_i =- 2J\sum_j{}^{'}  \psi_j     +g \vert \psi_i \vert^2 \psi_i- \mu  \psi_i,
\end{equation}
where the primed sum above is over $j$, which are nearest-neighbor lattice sites to $i$, and $J\propto 1/2m$. We have also included a chemical potential ($\mu$) explicitly. It is natural to expect that the long distance behavior of this lattice system is similar to that of the GP equation, when it describes a superfluid phase.  

Equivalently, Eq.~\ref{lattice_gpe} is the Heisenberg equation of motion of the second-quantized boson field $\hat{\psi}_i$, that is, $\textrm{i}\hbar\del_t\hat{\psi}=[\hat{\psi}, \hat{H}]$, obtained from an appropriate Hamiltonian. This is the so-called Bose-Hubbard model with a Hamiltonian (in second-quantized notation) \cite{fisher1989boson},
\begin{equation}\label{bose-hubbard}
    \hat{H}=-J\sum_{\langle ij\rangle}\left[\hat{\psi}_i^\dagger \hat{\psi}_j+\textrm{h.c.}\right]+\frac{g}{2}\sum_i\hat{\psi}_i^\dagger \hat{\psi}_i(\hat{\psi}^\dagger_i\hat{\psi}_i-1)-\mu\hat{\psi}_i^\dagger \hat{\psi}_i, 
\end{equation}
where h.c. denotes Hermitian conjugate, $\langle ij\rangle$ denotes a sum over nearest neighbors $i$ and $j$, and the bosonic fields obey the algebra, 
\begin{equation}
[\hat{\psi}_i,\hat{\psi}_j^\dagger]=\delta_{ij},\;\;\; [\hat{\psi}_i,\hat{\psi}_j]=0,\;\;\;\textrm{and}\;\;\; [\hat{\psi}^\dagger_i,\hat{\psi}_j^\dagger]=0.
\end{equation}
The lattice model above suggests new directions of exploration. A key advantage of the lattice formulation  is that it addresses possible phases that may occur in the strong coupling limit, where $g \gg J$, which are inacessible in the continuum. Consider again the Bose-Hubbard Hamiltonian, in Eq.~\ref{bose-hubbard}. When $g \gg J$, the lattice filling plays a crucial role in determining the ground states; when the boson density per site is an integer, the resulting ground state is known as a Mott insulator. This state has no low energy degrees of freedom and one  would not naturally expect hydrodynamic behavior in this regime [the grey region in Fig.~\ref{fig:many-body}(b)]. When $g/J\ll1$, the bosons achieve phase coherence and realize a superfluid. See Fig.~\ref{fig:many-body}(a) for a cartoon picture of the two phases. There must be a zero temperature phase transition separating the superfluid phase at $g/J \ll 1$ from the Mott insulating phase when $g/J \gg 1$. If this transition is continuous, as in this case, universal scaling laws as a function of temperature and $g/J$ occur, and such continuous $T=0$ phase transitions are known as quantum critical points \cite{sachdev1999quantum}. 

We have described in the section before the possible consequences for quantum turbulence in such a critical fluid. Furthermore, this question may be investigated as a function of temperature in this fluid, since the universal scaling of the quantum critical fluid persists within a range of finite temperature, in a region of parameter space known as the quantum critical fan (see Fig.~\ref{fig:many-body}(b) for a finite-temperature phase diagram of the two-dimensional Bose-Hubbard model). In two dimensions, the Wilson-Fisher critical point occurs as a transition at fixed commensurate density, say, of one boson per site. While  emergent momentum conservation, key to a hydrodynamic description, is ideally obtained at dilute densities in the lattice superfluid, we note that the critical fan realizes a scaling regime where fluctuations occur at all length scales, and the lattice constant is no longer important. Note that the classical finite temperature two-fluid model is not valid here, since there is no apparent differentiation between a 'super' and 'normal' fluid component. 

Since a hydrodynamical description of a many-body system is typically at a coarse-grained level, the participating degrees of freedom are often classical. Regardless, identification of the effective 'classical' degrees of freedom in a many-body quantum fluid may itself be a nontrivial task. For instance, near the superfluid-insulator transition, at low temperatures, we suspect that a \textit{quantum} two-fluid model may be an efficient description of turbulent hydrodynamics~\cite{BhattacharjeeHelbigRaghu}. Unlike the classical Landau two-fluid model describing a coupled super- and normal fluid, the quantum two-fluid description would be a theory of the coupled super- and \textit{vortex} fluids. In the former, thermally induced vortex motion is key, while in the latter, one must consider the quantum dynamics of the vortices. Both fluids (boson and vortex) can undergo turbulent hydrodynamics as long as the forcing amplitude is at least comparable to their energy gaps. Considering such a description from the superfluid side, one may interpret the strong quantum fluctuations in the superfluid induced near the critical point as being captured by a new 'classical' variable---the vortex degree of freedom. We highlight that in this regime even though the effective hydrodynamics is classical, it is classical in a new set of variables---one that is inaccessible from a mean-field description such as the GP equation. This thus requires a many-body quantum description of the phenomena, such as a boson-vortex duality~\cite{peskin1978mandelstam}. Thus even though the hydrodynamics may be classical, we consider it fit to call such a regime a 'many-body' regime of quantum turbulence.

Before discussing turbulence in a one-dimensional superfluid, we note that one may also examine the consequence of increasing the strength of interactions \textit{within} the single-particle formulation. This may be interpreted as marching towards the critical point starting from the superfluid, albeit ignoring quantum fluctuations within the classical formulation of the GP equation. In that case, as noted in Sec.~\ref{sec:quantum_fluids}, increasing $g$ is akin to increasing temperature in the corresponding classical fluid. This increases the speed of sound, and hence decreases the Mach number of the fluid, reducing the effects of compressibility. (The Mach number for the weakly-interacting superfluid may be given by $\textrm{Ma}_q=\bar{v}/c_s$, where $c_s=\sqrt{g\rho_0}$ and $\bar{v}$ is the average velocity.) Another consequence of this is that in the limit when $g\rightarrow \infty$, which is the Thomas-Fermi limit in the absence of a lattice, the vortex core size $\xi\rightarrow 0$. In that case, the inertial range extends up to the largest wavenumbers, increasing the width of wavenumber range over which a Kelvin wave cascade can be sustained. Effects beyond mean-field can be systematically included in the GP equation in the dilute limit using the Lee-Huang-Yang correction \cite{lee1957eigenvalues}; however, this simply renormalizes the speed of sound, without accounting for increased quantum fluctuations. Verifying this expectation more accurately in a numerical calculation of the GP equation will likely be of interest to the quantum turbulence community.

\subsection{Turbulence in a one dimensional quantum fluid}\label{one_d_problem}
We very briefly describe prospects for quantum turbulence with strong quantum fluctuations in one spatial dimension, since the general idea has been laid out in the previous subsection. Here, one may consider a lattice model, such as the Bose-Hubbard model again. The lack of long-range order in one dimension implies that the superfluid has power-law correlations even deep within the phase, and at long-distances behaves like a Luttinger liquid \cite{haldane2005luttinger}. The GP model fails to account for quantum fluctuations in such a fluid; e.g., the density and phase variables in a Luttinger liquid obey a commutation relation, given by,
\begin{equation}
    [\hat{\rho}(x),\hat{\theta}(x')]=\textrm{i}\delta(x-x'),
\end{equation}
which is not accounted for in the GP equation. At distances small compared to the phase coherence length, however, the commutation relation becomes unimportant, and the GP equation is often an adequate description. At weak driving, the sound waves (phonons modes) of such a one dimensional superfluid may undergo wave turbulence~\cite{nazarenko2011wave}. 

Another important difference in one dimension is that of the excitations, which are now phase-slips (the one-dimensional counterpart of vortices). While in the GP equation they are obtained as classical solutions, in one-dimension, especially at strong driving, it appears to be important to consider the quantum dynamics of these phase-slips, which can also lead to new cascade power laws. The quantum phase transition in the one-dimensional lattice model is in the Berezinskii-Kosterlitz-Thouless universality class \cite{berezinskii1972destruction, kosterlitz1973ordering}, and occurs through proliferation of vortices, which are phase-slips (or instantons) in space-time. This one dimensional quantum critical point has conformal invariance and the study of quantum turbulence at criticality may have some connections to the study of conformal field theories with quantum quenches~\cite{calabrese2016}. This also appears to be a fertile regime for many-body quantum effects in turbulent dynamics.

\section{Discussion}\label{discussion}

In recent times, questions pertaining to many-body effects in the dynamics of quantum systems are actively being investigated. As discussed above, these issues arise in the context of how closed quantum systems thermalize, but also its lack thereof. The latter has led to the discovery of interesting non-equilibrium phenomena such as many-body localization \cite{oganesyan2007localization, nandkishore2015many, abanin2019colloquium} and quantum scars \cite{bernien2017probing, abanin2019colloquium}. In this evolving landscape, we believe that a serious consideration of the problem of quantum turbulence in the many-body regime can also yield a greater understanding of non-equilibrium quantum many-body physics.    

On the experimental side, increasing developments in cold-atom condensates and quantum computing platforms have led to realizations of lattice models of bosons, such as the Bose-Hubbard and quantum XY models, which can access the quantum critical regime \cite{greiner2002quantum, bernien2017probing}. It would be of interest to determine driving protocols and dissipation mechanisms in such setups to study the problem of many-body quantum turbulence. In this regard, we also highlight clean superconducting thin films as a promising platform. In a two-dimensional superconductor, three dimensional electromagnetic fields only weakly lift the Nambu-Goldstone mode to the plasmon frequency, and hence at forcing amplitudes large compared to this energy scale, it should be possible to obtain turbulent cascades in its hydrodynamics.

Furthermore, the interaction of structured light with quantum matter opens new avenues for exploring turbulent dynamics in superconductors. In particular, shining light carrying spin angular momentum and orbital angular momentum onto a superconducting film enables the controlled excitation of superflow, vortices, and collective modes, including Higgs modes \cite{aeppli2025quantum}. This process, termed quantum printing, imprints the photonic quantum numbers directly into the superconducting fluid, offering a novel route to manipulate vortices and collective excitations. Unlike superfluid turbulence, a quantum mechanical description is required even for the forcing. This can  fundamentally alter the governing physics---for example, a steady influx of angular momentum into the fluid can alter the nature of the turbulent cascades, or induce non-equilibrium phase transitions, such as a light-induced Berezinskii–Kosterlitz–Thouless (BKT) transition \cite{yeh2025light}. A more thorough theoretical analysis of the underlying physics is wanted and may lead to a realization of turbulent cascades in a solid-state experiment.

We reviewed superfluid turbulence in Sec.~\ref{sec:quantum_fluids}. Our discussion here was limited to three dimensions; in lower dimensions (such as two), additional conservation laws can lead to a richer scaling law structure, and invert the direction of the cascade from the UV to IR. This is true for the classical fluid \cite{Kraichnan:PF1967_2D}, and may also occur in the Gross-Pitaevskii fluid, particularly when the vortices in the steady-state form like-signed clusters~\cite{reeves2013inverse}. In Sec.~\ref{many-body}, we provided concrete examples of fluids that can show turbulent dynamics with quantum many-body effects; such effects are particularly prominent in lower dimensions (such as one and two). We thus anticipate that conservation laws can play an important role in the nature of the direction of the cascades. We also think that in low dimensions, powerful numerical techniques, such as exact diagonalization \cite{lin1990exact} and the density matrix renormalization group (DMRG) \cite{white1992density} can shed some light.

In Sec.~3\ref{lattice_model}, we discussed the need for a quantum many-body description of the system to identify the emergent hydrodynamical variables that exhibit turbulence near a quantum critical point. The kinetic coefficients (such as viscosity) that enter such a fluid equation depend on the interactions between the degrees of freedom, and are renormalized by quantum fluctuations~\cite{muller2009graphene, kovtun2005viscosity}. At length-scales small compared to a mean-free path $\propto h/k_BT$, in the quantum critical fan, an emergent hydrodynamic description may cease to exist and a \textit{quantum} kinetic theory~\cite{kamenev2023field} may be required to describe its dynamics. We think that a study of the hydrodynamic equations coupled to a noisy term~\cite{Hohenberg:RMP1977, kamenev2023field} (where the noise represents quantum fluctuations) may also be useful to uncover the associated turbulent dynamics.

The problem of turbulence of systems with strong quantum fluctuations is inherently non-perturbative in character. Thus to make progress, one must invoke a variety of approaches, ranging from seeking dual representations for hydrodynamics \cite{peskin1978mandelstam, BhattacharjeeHelbigRaghu}, to using computational techniques in low dimensions as outlined above. We also believe that the framework of 'non-thermal (or dynamical) fixed points'~\cite{karl2017strongly, schmied2019non} could help classify the non-equilibrium scaling regimes obtained as a result of turbulent flow of quantum fluids, especially in the absence of a drive. Nevertheless, a possibly analytically tractable limit might exist, which involves turbulent hydrodynamics near a quantum critical point in \textit{three} spatial dimensions.  In equilibrium, the superfluid-insulator quantum critical point in three spatial dimensions, when enriched by particle-hole symmetry enjoys a description in terms of a Gaussian fixed point.  Correlation functions can be extracted from mean-field theory up to logarithmic corrections to scaling. One may examine, using techniques such as the dynamical renormalization group, whether hydrodynamics in this turbulent regime continues to remain weakly coupled and compute the corrections to classical scaling.     

We believe that we have identified a fertile ground for novel explorations of quantum many-body dynamics. Much work remains to be done to answer some of the questions posed in this review, and we believe that these questions can be actively pursued by the physics community in the years to come.

\dataccess{This article has no additional data.}
\funding{MKV was supported by J. C. Bose Fellowship (SERB/PHY/2023488). AB was supported by  the U.S. Department of
Energy,  Office of Basic Energy Sciences
under Award No. DE-SC-0025580. SB and SR are supported in part by the US Department of Energy, Office of Basic Energy Sciences, Division
of Materials Sciences and Engineering, under Contract
No. DE-AC02-76SF00515.}
\conflict{The authors have no conflicts of interest to declare.}

\ack{The authors thank  V. Khemani, A. Lakshminarayan, R. Laughlin, Y. Li, S. Sondhi and T.-T. Yeh for valuable  discussions. We also thank T. Helbig, N. O'Dea, J. Yu and the anonymous referees for a careful reading of the manuscript and useful suggestions. Part of this work was done in the Center for Turbulence Research, Stanford University, where MKV was a Visiting Senior Fellow. SR and MKV also thank the Field Theory and Turbulence Discussion Meeting held at the International Center for Theoretical Sciences, Bengaluru, India for stimulating some of the ideas discussed in this article.}

\bibliographystyle{RS}
\bibliography{journal.bib}

\end{document}